\documentclass[12pt,a4paper]{article}
\usepackage{amsmath,amsfonts,amsthm,graphicx,lscape}
\usepackage[dvips]{psfrag}
\usepackage{cite}

\usepackage[normalem]{ulem}

\usepackage{textcomp}
\usepackage{amsmath,amsthm,amssymb,mathrsfs,cases}
\usepackage{amsfonts,graphicx,lscape}

\usepackage{accents}
\makeatletter
\def\widebar{\accentset{{\cc@style\underline{\mskip10mu}}}}
\makeatother

\numberwithin{equation}{section}

\def\beqa{\begin{eqnarray}}
\def\enqa{\end{eqnarray}}
\def\beq{\begin{equation}}
\def\enq{\end{equation}}

\begin{document}
\title{
On an 
integrable 
discretization
of 
the massive Thirring model in non-characteristic
coordinates}
\author{Takayuki \textsc{Tsuchida}
}
\maketitle
\begin{abstract} 
We propose 
the Lax-pair representation for 
an integrable semi-discretization (discretization of the spatial variable) of 
the massive Thirring model in non-characteristic 
(in between light-cone and laboratory) coordinates 
and present its $N$-soliton solution. 
\end{abstract}
%


\newpage
\section{Introduction}

A 
long-standing problem 
in the theory of classical integrable systems is 
to construct 
integrable 
discrete analogs 
of the 
massive Thirring model~\cite{Thirring58}  
in laboratory coordinates. 
As early as in 1983, 
Nijhoff, Capel 
and Quispel~\cite{NCQ83} 
considered 
a closely related (but simpler) problem;  
on the basis of their previous 
work~\cite{NCQL},
they obtained 
integrable discretizations of 
the massive Thirring model 
rewritten in light-cone 
(or characteristic) 
coordinates. 
Note that their discretizations 
appear to be 
rather 
complicated and not easily 
recognizable 
as 
discrete analogs of the massive Thirring model. 
More recently, we proposed 
simpler integrable discretizations of 
the massive Thirring model in 
light-cone 
coordinates~\cite{Me2015,Me2024}. 
However, these 
results do not 
translate directly 
to 
proper 
integrable discretizations 
of the massive Thirring model in laboratory coordinates.

We can
construct 
an integrable discretization 
of the massive Thirring model 
in laboratory coordinates, 
which, however,  
involves undesired auxiliary dependent variables 
and look awfully cumbersome. 
Thus, we alternatively 
address a more tractable problem:\ 
to construct 
an aesthetically acceptable integrable discretization 
of the massive Thirring model in 
non-characteristic and non-laboratory 
(i.e., in between light-cone and laboratory) coordinates. 
An integrable semi-discretization 
(discretization of the spatial variable) 
of the massive Thirring model in 
non-characteristic and non-laboratory 
coordinates 
was recently 
proposed by 
Joshi and Pelinovsky~\cite{Peli_Joshi} and 
its soliton solutions were derived 
by 
Xu and Pelinovsky~\cite{Pelinovsky2019}.\footnote{
Joshi and Pelinovsky~\cite{Peli_Joshi} 
applied a coordinate transformation 
after taking a 
continuous limit and claimed that 
they derived an integrable semi-discretization 
of the massive Thirring model in laboratory coordinates. 
However, it is more natural to understand 
that they 
obtained an integrable semi-discretization 
of the massive Thirring model in 
non-characteristic and non-laboratory 
coordinates. 
}
Their results~\cite{Peli_Joshi,Pelinovsky2019} appear to be interesting and meaningful,  
but 
their semi-discretization 
involves 
an 
undesired 
auxiliary dependent variable. 

In this paper, we 
propose 
an 
integrable semi-discretization 
of the massive Thirring model in non-characteristic 
and non-laboratory 
coordinates, 
which can be expressed explicitly 
without using any 
auxiliary dependent variables; 
the equations of motion that we discretize (see~(\ref{cMTM0}))
are 
essentially 
the same as those considered by Pelinovsky and coworkers~\cite{Peli_Joshi,Pelinovsky2019}. 
To achieve our goal, 
we note 
a close relationship 
between 
an integrable discrete nonlinear Schr\"odinger (Ablowitz--Ladik~\cite{AL1,AL76}) hierarchy 
and an integrable derivative nonlinear Schr\"odinger (Chen--Lee--Liu~\cite{CLL}) hierarchy 
that contains 
the massive Thirring model 
as a particular flow~\cite{NCQL}. 
More specifically, 
the 
Lax-pair
representation~\cite{Lax} for the massive Thirring model~\cite{Kuz} 
and (the temporal part of) the 
Lax-pair representation~\cite{Lax} for the  Ablowitz--Ladik lattice~\cite{AL1,AL76} 
share a striking resemblance 
in 
their 
dependence on the spectral parameter. 
This resemblance is not an accidental coincidence 
(cf.~\cite{Vek98,Vek02}); 
note also that (the normalized form of)
the Ablowitz--Ladik spectral problem~\cite{AL1,AL76}  
appears 
as a B\"acklund--Darboux
transformation for the 
Chen--Lee--Liu hierarchy~\cite{Getmanov87,Getmanov93}. 
To obtain 
an integrable 
semi-discretization 
of the massive Thirring model, 
we 
compose its Lax-pair representation~\cite{Lax}  
by combining the temporal part of the 
Lax-pair representation for the Ablowitz--Ladik lattice~\cite{AL1,AL76} 
and 
that for a proper 
discrete-time flow of the Ablowitz--Ladik 
hierarchy~\cite{AL76,AL77} 
(see also \cite{Chiu77}).

This paper is organized as follows. 
In section~2, 
by borrowing an idea from the 
work of Ablowitz and Ladik~\cite{AL1,AL76,AL77}, 
we 
construct 
the Lax-pair representation 
for an integrable 
semi-discretization 
of the massive Thirring model 
in non-characteristic and non-laboratory coordinates; we
also use our own method~\cite{2010JPA} 
to eliminate 
an 
undesired 
auxiliary dependent 
variable that appears 
in the 
discrete-time 
flow of the Ablowitz--Ladik hierarchy~\cite{AL76,AL77}. 
In section~3, 
we 
construct the $N$-soliton 
solution 
of the 
semi-discrete massive Thirring model 
using a method based on the inverse scattering transform. 
Section~4 is devoted to conclusions. 
In the appendix, 
we show how 
the semi-discrete Lax pair in section~2 
can be related to the Lax pair 
for the continuous massive Thirring model in light-cone coordinates 
and the associated binary B\"acklund--Darboux transformation.

\section{Semi-discrete massive 
Thirring model}

In this section, we 
combine 
the 
temporal 
Lax matrices 
for discrete-time and continuous-time flows of 
the Ablowitz--Ladik hierarchy~\cite{AL1,AL76,AL77} 
to compose 
the Lax-pair representation for 
an integrable 
semi-discretization of the massive Thirring model 
in non-characteristic and non-laboratory coordinates. 
We also discuss some properties of the semi-discrete massive Thirring model. 

\subsection
{Lax pair}
\label{section2.1}

We consider the pair of semi-discrete 
linear equations: 
\begin{subequations}
\label{line}
\begin{align}
& \Psi_{n+1} = L_n (\zeta)
\Psi_n, 
\label{line_s}
\\
& \Psi_{n,t} = M_n (\zeta)\Psi_n.
\label{line_t}
\end{align}
\end{subequations}
Here, \mbox{$n \in \mathbb{Z}$} is the discrete spatial variable, 
the subscript $t$ denotes the 
differentiation
with respect to 
the continuous time variable $t$ and $\zeta$ is a constant 
spectral parameter; 
$\Psi_n$ is a 
column vector 
and the square 
matrices $L_n$
and $M_n$ 
constitute the Lax pair~\cite{Lax}. 
The compatibility condition of the overdetermined linear equations 
(\ref{line}) 
is given by the matrix equation~\cite{AL1,Kako,AL76,Ize81}:
\begin{equation}
 L_{n,t} = M_{n+1}L_n - L_n M_n,
\label{Lax_eq00}
\end{equation}
which is
(a semi-discrete version of) the zero-curvature
condition. 
%
The zero-curvature condition (\ref{Lax_eq00}) 
implies the simple 
conservation law:
\begin{equation}
\frac{\partial}{\partial t} \log (\det L_n) 
= 
\mathrm{tr} \hspace{1pt} M_{n+1} - \mathrm{tr} \hspace{1pt} M_{n}. 
\label{sd-cons}
\end{equation}
For a proper 
choice of $L_n$, 
\mbox{$\det L_n
$} 
is either 
a 
time-independent 
function of $\zeta$ 
or the exponential of 
a $\zeta$-independent 
conserved density 
%
multiplied by 
a
time-independent 
factor~\cite{2010JPA}. 

In this paper, we consider the 
discrete spatial Lax matrix: 
\begin{align}
L_n (\zeta)= 
\left[
\begin{array}{cc}
 \Lambda_n & 0 \\
 0 & 1 \\
\end{array}
\right] - \frac{\mathrm{i}\varDelta}{2}
 \left[
\begin{array}{cc}
 \zeta^2 & \zeta (\Lambda_n q_{n+1} + q_n) \\
 \zeta (r_{n+1} + \Lambda_n r_n) & -\zeta^2 \Lambda_n \\
\end{array}
\right].
\label{sdMTM_L}
\end{align}
Here, $\mathrm{i}$ is the imaginary unit, 
$\varDelta$ is a (typically small but nonzero) lattice parameter, 
$q_n$ and $r_n$ are dependent variables 
and $\Lambda_n$
is an auxiliary function to be determined later. 
We assume the boundary conditions: 
\begin{equation}
\lim_{n \to \pm \infty}
q_n =
\lim_{n \to \pm \infty}
r_n=0, \hspace{5mm}
\lim_{n \to - \infty} \Lambda_n =1.
\label{qrL_bound}
\end{equation}
We associate 
(\ref{sdMTM_L}) 
with 
the temporal Lax matrix: 
\begin{align}
M_n (\zeta) = 
\mathrm{i} a\left[
\begin{array}{cc}
 -q_n r_n & \zeta  q_n \\
 \zeta r_n & - \zeta^2 \\
\end{array}
\right]
+\mathrm{i} b
\left[
\begin{array}{cc}
 -\frac{1}{\zeta^2} & \frac{1}{\zeta} u_n \\[1mm]
 \frac{1}{\zeta} v_n & -v_n u_n \\
\end{array}
\right],
\label{sdMTM_M}
\end{align}
where 
$u_n$ and $v_n$ are dependent variables and 
the 
constant parameters 
$a$ and $b$ 
satisfy the condition \mbox{$(a,b) \neq (0,0)$} 
for 
the nontriviality of the time evolution. 

Substituting (\ref{sdMTM_L}) and 
(\ref{sdMTM_M}) into the zero-curvature condition (\ref{Lax_eq00}),  
we obtain the following 
differential-difference
system: 
\begin{subequations}
\label{system00}
\begin{align}
& \mathrm{i} (\Lambda_n q_{n+1} + q_n)_t + a \left\{ 
 \frac{2\mathrm{i}}{\varDelta} (q_{n+1} - \Lambda_n q_n) -q_{n+1} r_{n+1} (\Lambda_n q_{n+1} + q_n) \right\}
\nonumber \\
& - b \left\{ (\Lambda_n u_{n+1} + u_n) -u_n v_n (\Lambda_n q_{n+1} + q_n) \right\}=0, 
\label{qq_t}
\\[1mm]
& \mathrm{i} (r_{n+1} + \Lambda_n r_n)_t + a \left\{ \frac{2\mathrm{i}}{\varDelta}  (\Lambda_n r_{n+1} - r_n) + q_n r_n  (r_{n+1} + \Lambda_n r_n) \right\}
\nonumber \\
& - b \left\{ -(v_{n+1} + \Lambda_n v_n) 
	+ u_{n+1} v_{n+1} (r_{n+1} + \Lambda_n r_n) \right\}=0,
\label{rr_t}
\\[1mm]
&  \frac{2\mathrm{i}}{\varDelta} (u_{n+1}-\Lambda_n u_n) - 
(\Lambda_n q_{n+1} + q_n ) =0 \;\,\, \mathrm{if} \; b\neq 0, 
\label{uuqq}
\\[1mm]
&  \frac{2\mathrm{i}}{\varDelta} (\Lambda_n v_{n+1}-v_n) + 
(r_{n+1} + \Lambda_n r_n ) =0 \;\,\, \mathrm{if} \; b\neq 0, 
\label{vvrr}
\\[1mm]
& \mathrm{i} \Lambda_{n,t} = a \Lambda_n (q_{n+1} r_{n+1} - q_n r_n) + b \Lambda_n (u_{n+1} v_{n+1} - u_n v_n).
\label{Lambda_t}
\end{align}
\end{subequations}
In the special case 
\mbox{$b=0$}, 
the dependent variables $u_n$ and $v_n$ do not appear in 
(\ref{sdMTM_M}), so 
we do not have (\ref{uuqq}) and (\ref{vvrr}). 

For the 
Lax matrix (\ref{sdMTM_L}), 
we 
compute its determinant as 
\begin{align}
& \mathrm{det} L_n (\zeta) 
 = \left(1 +\frac{\varDelta^2}{4} \zeta^4 \right ) \Lambda_n 
\nonumber \\
&	+ \frac{\mathrm{i}\varDelta}{2} \zeta^2 \left[ \left( 1- \frac{\mathrm{i}\varDelta}{2} q_{n+1} r_n \right) \Lambda_n^2 
	- \frac{\mathrm{i}\varDelta}{2} 
	(q_{n+1} r_{n+1} + q_n r_n) \Lambda_n  - \left( 1 + \frac{\mathrm{i}\varDelta}{2} q_n r_{n+1} \right) \right]. 
\nonumber
\end{align}
%
Thus, the 
conservation law (\ref{sd-cons}) implies
not only (\ref{Lambda_t}) 
but also the following nontrivial 
relation: 
\begin{align}
& \mathrm{i} \left[ \left( 1- \frac{\mathrm{i}\varDelta}{2} q_{n+1} r_n \right) \Lambda_n^2 - \frac{\mathrm{i}\varDelta}{2} 
	(q_{n+1} r_{n+1} + q_n r_n) \Lambda_n  - \left( 1 + \frac{\mathrm{i}\varDelta}{2} q_n r_{n+1} \right) \right]_t
\nonumber \\
&=	\left[ \left( 1- \frac{\mathrm{i}\varDelta}{2} q_{n+1} r_n \right) \Lambda_n^2 - \frac{\mathrm{i}\varDelta}{2} 
	(q_{n+1} r_{n+1} + q_n r_n) \Lambda_n  - \left( 1 + \frac{\mathrm{i}\varDelta}{2} q_n r_{n+1} \right) \right]
\nonumber \\
& \times 
\left[ a (q_{n+1} r_{n+1} - q_n r_n) + b (u_{n+1} v_{n+1} - u_n v_n) \right]. 
\label{nont_con}
\end{align}
Note that 
this 
relation is 
a direct consequence of 
(\ref{system00}). 
Comparing (\ref{nont_con}) with (\ref{Lambda_t}), 
we have 
\begin{align}
\left\{
\frac{1}{\Lambda_n} 
\left[ \left( 1- \frac{\mathrm{i}\varDelta}{2} q_{n+1} r_n \right) \Lambda_n^2 - \frac{\mathrm{i}\varDelta}{2} 
	(q_{n+1} r_{n+1} + q_n r_n) \Lambda_n  - \left( 1 + \frac{\mathrm{i}\varDelta}{2} q_n r_{n+1} \right) \right]
\right\}_t =0. 
\nonumber 
\end{align}
In view of the boundary conditions (\ref{qrL_bound}), 
we 
set 
\begin{align}
\left( 1- \frac{\mathrm{i}\varDelta}{2} q_{n+1} r_n \right) \Lambda_n^2 - \frac{\mathrm{i}\varDelta}{2} 
	(q_{n+1} r_{n+1} + q_n r_n) \Lambda_n  - \left( 1 + \frac{\mathrm{i}\varDelta}{2} q_n r_{n+1} \right) =0, 
\label{L_quad}
\end{align}
and 
choose the solution of this quadratic equation as 
\begin{align}
& \Lambda_n  (q_{n+1}, r_{n+1}, q_{n}, r_{n}) 
\nonumber \\
& = \frac{\sqrt{\left( 1- \frac{\mathrm{i}\varDelta}{2} q_{n+1} r_n \right) \left( 1 + \frac{\mathrm{i}\varDelta}{2} q_n r_{n+1} \right)
	- \frac{\varDelta^2}{16} (q_{n+1} r_{n+1} + q_n r_n)^2} + \frac{\mathrm{i}\varDelta}{4} (q_{n+1} r_{n+1} + q_n r_n)}
{1- \frac{\mathrm{i}\varDelta}{2} q_{n+1} r_n}. 
\label{L_exp1}
\end{align}
The square root of the 
complex 
function in 
(\ref{L_exp1}) is defined as the Maclaurin series in $\varDelta$, so 
(\ref{L_exp1})
admits the following expansion 
for a small value of the lattice parameter $\varDelta$: 
\begin{align}
\Lambda_n  (q_{n+1}, r_{n+1}, q_{n}, r_{n}) 
= 1 + \frac{\mathrm{i}\varDelta}{4} (q_{n+1} + q_n) (r_{n+1} + r_n) + \mathcal{O}(\varDelta^2). 
\label{L_exp2}
\end{align}
Note that the explicit expression (\ref{L_exp1}) 
for $\Lambda_n$ 
implies the quadratic equation (\ref{L_quad}), which can be rewritten as 
\begin{align}
\left( \Lambda_n^2-1 \right) - \frac{\mathrm{i}\varDelta}{2} (\Lambda_n q_{n+1} + q_n) (r_{n+1} + \Lambda_n r_n) = 0. 
\label{L_quadra}
\end{align}
By differentiating (\ref{L_quadra}) with respect to 
$t$ 
and using 
(\ref{qq_t})--(\ref{vvrr}) and (\ref{L_quadra}), 
we can derive the relation 
(\ref{Lambda_t}). 
Thus, if we adopt (\ref{L_exp1}) as the definition of the auxiliary function $\Lambda_n$, 
(\ref{Lambda_t}) is no longer necessary and can be discarded.

\subsection{Complex conjugation reduction}
\label{subsec2.2}

If the parameters appearing in (\ref{system00}) 
are all real, i.e., 
\mbox{$\varDelta, a, b \in \mathbb{R}$}, 
then we can impose the complex conjugation reduction: 
\begin{equation}
r_n = q_n^\ast, \hspace{5mm} v_n=u_n^\ast, \hspace{5mm} n \in \mathbb{Z}, 
\label{rq_vu}
\end{equation}
where the asterisk denotes the complex conjugate. 
Indeed, by imposing this reduction, 
(\ref{Lambda_t}) (or (\ref{L_quad}) or (\ref{L_exp1})) 
implies the relation 
\mbox{$\Lambda_n^\ast = \Lambda_n^{-1}$}. 
Using (\ref{Lambda_t}), 
we can rewrite (\ref{rr_t}) 
as 
\begin{align}
& \mathrm{i} \left( \Lambda_n^{-1} r_{n+1} + r_n \right)_t + a \left\{ \frac{2\mathrm{i}}{\varDelta}  \left( r_{n+1} - \Lambda_n^{-1} r_n \right) 
	+ q_{n+1} r_{n+1}  \left( \Lambda_n^{-1}r_{n+1} + r_n \right) \right\}
\nonumber \\
& - b \left\{ -\left( \Lambda_n^{-1} v_{n+1} + v_n \right) 
	+ u_{n} v_{n} \left( \Lambda_n^{-1} r_{n+1} + r_n \right) \right\}=0;
\nonumber
\end{align}
we also rewrite (\ref{vvrr}) as 
\begin{align}
&  \frac{2\mathrm{i}}{\varDelta} \left( v_{n+1}- \Lambda_n^{-1} v_n \right) + 
\left( \Lambda_n^{-1} r_{n+1} + r_n \right) =0.
\nonumber
\end{align}
These two relations 
can be obtained 
by taking the complex conjugate of (\ref{qq_t}) and (\ref{uuqq}), respectively. 
Thus, 
we 
arrive at the reduced semi-discrete system: 
\begin{subequations}
\label{sd_MTMeqs}
\begin{align}
& \mathrm{i} (\Lambda_n q_{n+1} + q_n)_t + a \left\{ 
 \frac{2\mathrm{i}}{\varDelta} (q_{n+1} - \Lambda_n q_n) -|q_{n+1}|^2 (\Lambda_n q_{n+1} + q_n) \right\}
\nonumber \\
& - b \left\{ (\Lambda_n u_{n+1} + u_n) -|u_n|^2 (\Lambda_n q_{n+1} + q_n) \right\}=0, 
\\[2mm]
&  \frac{2\mathrm{i}}{\varDelta} (u_{n+1}-\Lambda_n u_n) - 
(\Lambda_n q_{n+1} + q_n ) =0 \;\,\, \mathrm{if} \; b\neq 0,  
\\[2mm]
& 
\Lambda_n  
:= \frac{\sqrt{\left| 1- \frac{\mathrm{i}\varDelta}{2} q_{n+1} q_n^\ast \right|^2 
	- \frac{\varDelta^2}{16} \left( |q_{n+1}|^2 + |q_n|^2 \right)^2} + \frac{\mathrm{i}\varDelta}{4} \left( |q_{n+1}|^2 + |q_n|^2 \right)}
{1- \frac{\mathrm{i}\varDelta}{2} q_{n+1} q_n^\ast}. 
\end{align}
\end{subequations}

\subsection{Continuous limit}

To see that the semi-discrete system (\ref{sd_MTMeqs}) provides 
an integrable semi-discretization of the massive Thirring model, we note that 
the auxiliary function $\Lambda_n$ can be expanded as 
(cf.~(\ref{L_exp2}) and (\ref{rq_vu}))
\begin{align}
\Lambda_n 
= 1 + \frac{\mathrm{i}\varDelta}{4} \left| q_{n+1} + q_n \right|^2 + \mathcal{O}(\varDelta^2). 
\nonumber 
\end{align}
Thus, in the continuous limit \mbox{$\varDelta \to 0$}, 
(\ref{sd_MTMeqs}) indeed reduces 
to the 
massive Thirring model in 
non-characteristic and non-laboratory 
coordinates: 
\begin{equation} 
\label{cMTM0}
\left\{ 
\begin{split}
& \mathrm{i} ( q_{t}+ a q_x ) - b \left( u - |u|^2q \right) =0, 
\\[0.5mm]
& \mathrm{i} u_{x}- q + |q|^2 u =0,
\end{split} 
\right. 
\end{equation}
where $a$ and \mbox{$b (\neq 0)$} are real constants. 

The Lax-pair representation for 
(\ref{cMTM0}) 
is given by (cf.~\cite{Kuz,KN2,Morris79,KaMoIno,GIK})
\begin{align}
& 
\left[
\begin{array}{c}
 \Psi_{1} \\
 \Psi_{2} \\
\end{array}
\right]_{x}
= 
-\mathrm{i}
\left[
\begin{array}{cc}
 \frac{1}{2}\zeta^2 - |q|^2 & \zeta  q \\
 \zeta q^\ast & - \frac{1}{2}\zeta^2 \\
\end{array}
\right]
\left[
\begin{array}{c}
 \Psi_{1} \\
 \Psi_{2} \\
\end{array}
\right],
\nonumber \\[2mm]
& 
\left[
\begin{array}{c}
 \Psi_{1} \\
 \Psi_{2} \\
\end{array}
\right]_{t}
= 
\left\{\mathrm{i} a 
\left[
\begin{array}{cc}
 -|q|^2 & \zeta  q \\
 \zeta q^\ast & - \zeta^2 \\
\end{array}
\right]
+\mathrm{i}b
\left[
\begin{array}{cc}
 -\frac{1}{\zeta^2} & \frac{1}{\zeta} u \\[1mm]
 \frac{1}{\zeta} u^\ast & 
  - |u|^2 \\
\end{array}
\right] \right\}
\left[
\begin{array}{c}
 \Psi_{1} \\
 \Psi_{2} \\
\end{array}
\right],
\nonumber
\end{align}
which is obtained from (\ref{line})
with (\ref{sdMTM_L}) and (\ref{sdMTM_M}) 
by imposing the complex conjugation reduction 
and taking the continuous limit 
\mbox{$\varDelta \to 0$}.

\subsection{Decomposition of the semi-discrete Lax pair}

We would like to solve 
the semi-discrete massive Thirring model (\ref{sd_MTMeqs}) using 
a method based on 
the inverse scattering transform 
and construct its $N$-soliton solution. 
However, 
the discrete spectral problem (\ref{line_s}) with 
(\ref{sdMTM_L}) and (\ref{L_exp1}) 
is not amenable to the inverse scattering transform 
directly. 
Thus, we conceptually swap the roles of the spatial 
and temporal variables 
and 
consider the 
isospectral evolution equation (\ref{line_t}) 
with (\ref{sdMTM_M}) as the 
`spatial' spectral problem 
while considering (\ref{line_s}) with (\ref{sdMTM_L}) and (\ref{L_exp1}) as a discrete `time-evolution' equation.  

%
%
For the time being, 
we do not 
take into account the complex conjugation reduction 
discussed in subsection~\ref{subsec2.2}
or the explicit expression (\ref{L_exp1}) for the auxiliary function $\Lambda_n$, 
and consider the original 
system 
(\ref{system00}). 
It is in principle possible to apply the inverse scattering method to 
the spectral problem 
(\ref{line_t}) with (\ref{sdMTM_M}) directly, but this is still a 
laborious 
task~\cite{Kuz,KaMoIno}. 
Thus, we further decompose (\ref{line_t}) with (\ref{sdMTM_M}) 
into 
two linear problems 
as follows: 
\begin{subequations}
\label{MTM_t12}
\begin{align}
& 
\left[
\begin{array}{c}
 \Psi_{1,n} \\
 \Psi_{2,n} \\
\end{array}
\right]_{t_1}
= \mathrm{i}
\left[
\begin{array}{cc}
 -q_n r_n & \zeta  q_n \\
 \zeta r_n & - \zeta^2 \\
\end{array}
\right]
\left[
\begin{array}{c}
 \Psi_{1,n} \\
 \Psi_{2,n} \\
\end{array}
\right],
\label{U_t1}
\\[2mm]
& 
\left[
\begin{array}{c}
 \Psi_{1,n} \\
 \Psi_{2,n} \\
\end{array}
\right]_{t_2}
= \mathrm{i}
\left[
\begin{array}{cc}
 -\frac{1}{\zeta^2} & \frac{1}{\zeta} u_n \\[1mm]
 \frac{1}{\zeta} v_n & -v_n u_n \\
\end{array}
\right]
\left[
\begin{array}{c}
 \Psi_{1,n} \\
 \Psi_{2,n} \\
\end{array}
\right],
\label{U_t2}
\end{align}
\end{subequations}
where the decomposition means that the original time evolution is 
a linear combination 
\mbox{$\partial_t = a \partial_{t_1} + b \partial_{t_2}$}. 
In the next section, 
we take 
(\ref{U_t2}) as the main `spatial' spectral problem 
and regard (\ref{U_t1}) as an isospectral 
time-evolution 
equation associated with it. 

As 
shown in subsection~\ref{section2.1}, 
the compatibility condition of 
the overdetermined linear equations 
(\ref{line_s}) with (\ref{sdMTM_L}) and (\ref{U_t1}) 
is equivalent to the system: 
\begin{subequations}
\label{system01}
\begin{align}
& \mathrm{i} (\Lambda_n q_{n+1} + q_n)_{t_1} + 
 \frac{2\mathrm{i}}{\varDelta} (q_{n+1} - \Lambda_n q_n) -q_{n+1} r_{n+1} (\Lambda_n q_{n+1} + q_n) =0, 
\label{qq_t1}
\\[1mm]
& \mathrm{i} (r_{n+1} + \Lambda_n r_n)_{t_1} + \frac{2\mathrm{i}}{\varDelta}  (\Lambda_n r_{n+1} - r_n) + q_n r_n  (r_{n+1} + \Lambda_n r_n) =0, 
\label{rr_t1}
\\[1mm]
& \mathrm{i} \Lambda_{n,t_1} = \Lambda_n (q_{n+1} r_{n+1} - q_n r_n),
\label{Lambda_t1}
\end{align}
\end{subequations}
and the compatibility condition of 
the overdetermined linear 
equations (\ref{line_s}) with (\ref{sdMTM_L}) and (\ref{U_t2}) 
is equivalent to the system: 
\begin{subequations}
\label{system02}
\begin{align}
& \mathrm{i} (\Lambda_n q_{n+1} + q_n)_{t_2}  - (\Lambda_n u_{n+1} + u_n) +u_n v_n (\Lambda_n q_{n+1} + q_n) =0, 
\label{qq_t2}
\\[1mm]
& \mathrm{i} (r_{n+1} + \Lambda_n r_n)_{t_2} +(v_{n+1} + \Lambda_n v_n) 
	- u_{n+1} v_{n+1} (r_{n+1} + \Lambda_n r_n) =0,
\label{rr_t2}
\\[1mm]
&  \frac{2\mathrm{i}}{\varDelta} (u_{n+1}-\Lambda_n u_n) - 
(\Lambda_n q_{n+1} + q_n ) =0, 
\label{uuqq2}
\\[1mm]
&  \frac{2\mathrm{i}}{\varDelta} (\Lambda_n v_{n+1}-v_n) + 
(r_{n+1} + \Lambda_n r_n ) =0,
\label{vvrr2}
\\[1mm]
& \mathrm{i} \Lambda_{n,t_2} = \Lambda_n (u_{n+1} v_{n+1} - u_n v_n).
\label{Lambda_t2}
\end{align}
\end{subequations}

For 
each value of \mbox{$n \in \mathbb{Z}$}, 
the compatibility condition of 
(\ref{U_t1}) and (\ref{U_t2})
is equivalent to 
(the nonreduced form of) 
the massive Thirring model in light-cone 
coordinates~\cite{KN2,Morris79,GIK}: 
\begin{equation} 
\label{c-MTM1}
\left\{ 
\begin{split}
& \mathrm{i} q_{n,t_2} - u_n + q_n v_n u_n =0, 
\\[0.5mm]
& \mathrm{i} r_{n,t_2} + v_n - v_n u_n r_n =0,
\\[0.5mm]
& \mathrm{i} u_{n,t_1} + q_n - q_n r_n u_n =0, 
\\[0.5mm]
& \mathrm{i} v_{n,t_1} - r_n + v_n q_n r_n =0.
\end{split} 
\right. 
\end{equation}

Using (\ref{system01})--(\ref{c-MTM1}), 
we can directly check 
the commutativity of the two time derivatives: 
\begin{align}
& \partial_{t_1} \partial_{t_2} (\Lambda_n q_{n+1} + q_n) = \partial_{t_2} \partial_{t_1} (\Lambda_n q_{n+1} + q_n),
\nonumber \\[2mm]
& \partial_{t_1} \partial_{t_2} (r_{n+1} + \Lambda_n r_n) = \partial_{t_2} \partial_{t_1} (r_{n+1} + \Lambda_n r_n),
\nonumber 
\end{align}
as well as \mbox{$\partial_{t_1} \partial_{t_2} \Lambda_n  = \partial_{t_2} \partial_{t_1} \Lambda_n$}. 
Thus, if $q_n$, $r_n$, $u_n$ and $v_n$ satisfy the 
equations of motion for the 
massive Thirring model 
(\ref{c-MTM1}), 
the two systems (\ref{system01}) and (\ref{system02}) 
are indeed compatible
and the original system (\ref{system00}) 
can be decomposed consistently into two 
fundamental 
systems (\ref{system01}) and (\ref{system02}). 
Moreover, 
by adopting 
the explicit expression (\ref{L_exp1}) for the auxiliary function $\Lambda_n$ 
(cf.~the discussion in subsection~\ref{section2.1}), 
we can discard the time evolution equations (\ref{Lambda_t1}) and (\ref{Lambda_t2}) for 
$\Lambda_n$. 

\section{Soliton solutions}
In our previous paper~\cite{TsuJMP10}, 
we 
presented a 
set of solution formulas for 
the 
massive Thirring model  in light-cone 
coordinates 
under decaying boundary conditions 
(see~(2.12) and (2.38)
therein). 
Its derivation is based on the inverse scattering transform 
and the underlying idea is implicitly 
stated 
in~\cite{TsuJMP11} 
(see~Proposition~A.1 
therein). 
By considering 
(\ref{U_t2}) as 
the 
`spatial' spectral problem and 
rescaling the variables in~\cite{TsuJMP10} 
appropriately, 
we obtain the 
set of 
solution formulas for the 
massive Thirring model 
(\ref{c-MTM1})
under decaying boundary conditions 
as \mbox{$t_2 \to + \infty$}: 
%
%
\begin{subequations}
\label{Gel1}
\begin{align}
u_n (t_2) &= K_n (t_2,t_2),
\label{u_solu}
\\[2pt]
v_n (t_2) &= \widebar{K}_n (t_2,t_2),
\\[2pt]
K_n (t_2, y) 
&=
 \widebar{F}_n (y) -\mathrm{i}
\int^\infty_{t_2} \mathrm{d} s_1 \int^\infty_{t_2} \mathrm{d} s_2\hspace{1pt} 
K_n (t_2, s_1) F_n (s_1+s_2-t_2) \frac{\partial \widebar{F}_n (s_2+y-t_2)}{\partial s_{2}},
\hspace{5mm} y \ge t_2,
\label{Gel1c}
\\
\widebar{K}_n(t_2, y) 
&=
 F_n (y) + \mathrm{i} \int^\infty_{t_2} \mathrm{d} s_1 \int^\infty_{t_2} \mathrm{d} s_2\hspace{1pt} 
\frac{\partial \widebar{K}_n (t_2, s_1)}{\partial s_1} 
\widebar{F}_n(s_1+s_2-t_2) F_n (s_2+y-t_2),
\hspace{5mm} y \ge t_2,
\label{Gel1d}
\\[2pt]
q_n (t_2) &=J_n(t_2, t_2),
\label{q_solu}
\\[2pt]
r_n (t_2) &=\widebar{J}_n(t_2, t_2),
\\[2pt]
J_n(t_2, y)
&= \mathrm{i} \int_y^\infty \mathrm{d} s \hspace{1pt} \widebar{F}_n (s)
 + \mathrm{i} \int^\infty_{t_2} \mathrm{d} s_1 \int^\infty_{t_2} \mathrm{d} s_2\hspace{1pt}
J_n (t_2, s_1) \frac{\partial F_n (s_1+s_2-t_2)}{\partial s_2} \widebar{F}_n (s_2+y-t_2),
\nonumber \\
&
\hspace{110mm} y \ge t_2,
\label{Gel1g}
\\[2pt]
\widebar{J}_n(t_2, y)
&=
 -\mathrm{i} \int_y^\infty \mathrm{d} s \hspace{1pt} F_n (s)
 + \mathrm{i} \int^\infty_{t_2} \mathrm{d} s_1 \int^\infty_{t_2} \mathrm{d} s_2
\hspace{1pt}
\frac{\partial \widebar{J}_n (t_2, s_1)}{\partial s_1}
\frac{\partial \widebar{F}_n(s_1+s_2-t_2)}{\partial s_2} 
\int^\infty_{s_2+y-t_2} \mathrm{d} s_3 \hspace{1pt}F_n (s_3),
\nonumber \\
&
\hspace{110mm} 
y \ge t_2.
\end{align}
\end{subequations}
Here, 
the bar does not 
denote the 
complex
conjugate at this stage; 
for 
brevity, 
the 
$t_1$-dependence of 
all 
the functions 
appearing 
in 
(\ref{Gel1})
is suppressed. 

The 
functions 
$\widebar{F}_n(t_2)$ and $F_n (t_2)$, which 
also depend on $t_1$, 
satisfy the 
linear uncoupled 
system of partial differential equations: 
\begin{align}
\frac{\partial^2 \widebar{F}_n}{\partial t_1 \partial t_2} = \widebar{F}_n, 
\hspace{5mm}
\frac{\partial^2 F_n}{\partial t_1 \partial t_2} = F_n,
\label{dispers1}
\end{align}
and decay rapidly as \mbox{$t_2\to + \infty$}. 
%
%
Moreover, 
by considering 
(\ref{line_s}) with (\ref{sdMTM_L}) and (\ref{L_exp1}) as a discrete `time-evolution' equation, 
we also have the linear uncoupled system of differential-difference equations: 
\begin{align}
\frac{2}{\varDelta}
\frac{\partial}{ \partial t_2} \left( \widebar{F}_{n+1} - \widebar{F}_{n} \right)
	+ \widebar{F}_{n+1} + \widebar{F}_{n}=0, 
\hspace{5mm}
\frac{2}{\varDelta}
\frac{\partial}{\partial t_2} \left( F_{n+1} -  F_{n} \right)
+ F_{n+1}+ F_{n}=0.
\label{dispers2}
\end{align}
In short, the functions $\widebar{F}_n$ and $F_n$ 
are required 
to satisfy the linear part of the equations obeyed by $u_n$ and $v_n$ (cf.~(\ref{system02}) and (\ref{c-MTM1})).

The complex conjugation reduction 
(\ref{rq_vu}) 
can be realized in formulas (\ref{Gel1}) by 
imposing the reduction condition \mbox{$\widebar{F}_n=F_n^\ast$}. 
To obtain the 
$N$-soliton solution,  we 
consider the reflectionless case in the inverse scattering formalism
and 
choose the common solution of the linear systems (\ref{dispers1}) and (\ref{dispers2}) as 
\begin{subequations}
\label{FbarF}
\begin{align}
& \widebar{F}_{n} (t_2) = \sum_{j=1}^N c_j \left( \frac{2\lambda_j - \varDelta}{2\lambda_j + \varDelta} \right)^n 
\mathrm{e}^{\frac{1}{\lambda_j}t_1+\lambda_j t_2}, 
\\[1mm]
& F_{n} (t_2) =\sum_{j=1}^N c_j^\ast \left( \frac{2\lambda_j^\ast - \varDelta}{2\lambda_j^\ast + \varDelta} \right)^n 
\mathrm{e}^{\frac{1}{\lambda_j^\ast}t_1+\lambda_j^\ast t_2}. 
\end{align}
\end{subequations}
Moreover, we set 
\begin{subequations}
\label{KJ_exp}
\begin{align}
& K_{n} (t_2,y) = \sum_{j=1}^N f_n^{(j)}
c_j \left( \frac{2\lambda_j - \varDelta}{2\lambda_j + \varDelta} \right)^n 
	\mathrm{e}^{\frac{1}{\lambda_j}t_1+\lambda_j y}, 
\label{K_exp}
\\[1mm]
& J_{n} (t_2,y) = \sum_{j=1}^N g_n^{(j)}
c_j \left( \frac{2\lambda_j - \varDelta}{2\lambda_j + \varDelta} \right)^n 
	\mathrm{e}^{\frac{1}{\lambda_j}t_1+\lambda_j y}. 
\label{J_exp}
\end{align} 
\end{subequations}
Here, $c_j$ and $\lambda_j$ 
are nonzero complex constants, 
\mbox{${\rm Re}\, \lambda_j < 0$} \mbox{$(j=1,2, \ldots, N)$}, 
\mbox{$\lambda_j \neq \lambda_k$} 
if 
\mbox{$j \neq k$}, 
\mbox{$\varDelta \in \mathbb{R}$} and $f_n^{(j)}$ and $g_n^{(j)}$ 
are 
$y$-independent 
functions that depend on $t_1$ and $t_2$. 

Substituting (\ref{FbarF}) and (\ref{KJ_exp}) 
into (\ref{Gel1c}) and (\ref{Gel1g}),  
performing 
the 
integration 
and 
noting 
the 
linear independence of \mbox{$\mathrm{e}^{\lambda_1 y}$}, \mbox{$\mathrm{e}^{\lambda_2 y}$}, 
$\ldots$, \mbox{$\mathrm{e}^{\lambda_N y}$}, 
we obtain the relations for determining the 
unknown functions $f_n^{(j)}$ and $g_n^{(j)}$: 
\begin{align}
& \sum_{l=1}^N f_n^{(l)} \left( \delta_{lj} + U_{lj} \right) =1, 
\nonumber 
\\[1mm]
& \sum_{l=1}^N g_n^{(l)} \left( \delta_{lj} -V_{lj} \right) = -\frac{\mathrm{i}}{\lambda_j}.  
\nonumber
\end{align}
Here, 
$\delta_{lj}$ is 
the Kronecker delta and $U_{lj}$ and $V_{lj}$ are defined as 
\begin{subequations}
\label{UV_lj}
\begin{align}
& U_{lj} := \sum_{k=1}^N \frac{\mathrm{i} \lambda_j c_l c_k^\ast}
	{\left( \lambda_l + \lambda_k^\ast \right) \left( \lambda_k^\ast + \lambda_j \right)} 
	\left( \frac{2\lambda_l - \varDelta}{2\lambda_l + \varDelta} \right)^n 
\left( \frac{2\lambda_k^\ast - \varDelta}{2\lambda_k^\ast + \varDelta} \right)^n 
\mathrm{e}^{\left( \frac{1}{\lambda_l} + \frac{1}{\lambda_k^\ast} \right) t_1 + \left(\lambda_l +\lambda_k^\ast \right) t_2}, 
\label{U_lj} \\[1mm]
& V_{lj} := \sum_{k=1}^N \frac{\mathrm{i} \lambda_k^\ast c_l c_k^\ast}
	{\left( \lambda_l + \lambda_k^\ast \right) \left( \lambda_k^\ast + \lambda_j \right)} 
	\left( \frac{2\lambda_l - \varDelta}{2\lambda_l + \varDelta} \right)^n 
\left( \frac{2\lambda_k^\ast - \varDelta}{2\lambda_k^\ast + \varDelta} \right)^n 
\mathrm{e}^{\left( \frac{1}{\lambda_l} + \frac{1}{\lambda_k^\ast} \right) t_1 + \left(\lambda_l +\lambda_k^\ast \right) t_2}.
\label{V_lj}
\end{align}
\end{subequations}
Thus,  
$f_n^{(j)}$ and $g_n^{(j)}$ 
are determined 
as
\begin{subequations}
\label{QR-soliton}
\begin{align}
\left[
\begin{array}{cccc}
\! f_n^{(1)} \! & \! f_n^{(2)} \! & \! \cdots \! & \! f_n^{(N)} \! 
\end{array}
\right] 
= \left[
\begin{array}{cccc}
\! 1 \! & \! 1 \! & \! \cdots \! & \! 1 \! 
\end{array}
\right] 
\left\{ I_N + 
\left[
\begin{array}{ccc}
 U_{11} & \cdots & U_{1 N} \\
 \vdots &  \ddots & \vdots \\
 U_{N 1} & \cdots & U_{N N}\\
\end{array}
\right] \right\}^{-1}, 
\label{Q-Nsol}
\\[1mm]
\left[
\begin{array}{cccc}
\! g_n^{(1)} \! & \! g_n^{(2)} \! & \! \cdots \! & \! g_n^{(N)} \! 
\end{array}
\right] 
= -\mathrm{i} \left[
\begin{array}{cccc}
\! \frac{1}{\lambda_1} \! & \! \frac{1}{\lambda_2} \! & \! \cdots \! & \! \frac{1}{\lambda_N} \! 
\end{array}
\right] 
\left\{ I_N - 
\left[
\begin{array}{ccc}
 V_{11} & \cdots & V_{1 N} \\
 \vdots &  \ddots & \vdots \\
 V_{N 1} & \cdots & V_{N N}\\
\end{array}
\right] \right\}^{-1}, 
\label{R-Nsol}
\end{align}
\end{subequations}
where $I_N$ is the identity matrix of size $N$. 

Using (\ref{u_solu}), (\ref{q_solu}), 
(\ref{KJ_exp}) and
(\ref{QR-soliton}), 
we obtain 
\begin{subequations}
\label{soliton_t12}
\begin{align}
u_n 
&= \sum_{j=1}^N f_n^{(j)}
c_j \left( \frac{2\lambda_j - \varDelta}{2\lambda_j + \varDelta} \right)^n 
	\mathrm{e}^{\frac{1}{\lambda_j}t_1+\lambda_j t_2}
\nonumber \\
& = \left[
\begin{array}{cccc}
\! 1 \! & \! 1 \! & \! \cdots \! & \! 1 \! 
\end{array}
\right] 
\left\{ I_N + 
\left[
\begin{array}{ccc}
 U_{11} & \cdots & U_{1 N} \\
 \vdots &  \ddots & \vdots \\
 U_{N 1} & \cdots & U_{N N}\\
\end{array}
\right] \right\}^{-1}
\left[
\begin{array}{c}
 c_1 \left( \frac{2\lambda_1 - \varDelta}{2\lambda_1 + \varDelta} \right)^n 
	\mathrm{e}^{\frac{1}{\lambda_1}t_1+\lambda_1 t_2} \\
\vdots \\
 c_N \left( \frac{2\lambda_N - \varDelta}{2\lambda_N + \varDelta} \right)^n 
	\mathrm{e}^{\frac{1}{\lambda_N}t_1+\lambda_N t_2} \\
\end{array}
\right], 
\label{}
\\[2mm]
q_n 
&=\sum_{j=1}^N g_n^{(j)}
c_j \left( \frac{2\lambda_j - \varDelta}{2\lambda_j + \varDelta} \right)^n 
	\mathrm{e}^{\frac{1}{\lambda_j}t_1+\lambda_j t_2}
\nonumber \\
& = -\mathrm{i} \left[
\begin{array}{cccc}
\! \frac{1}{\lambda_1} \! & \! \frac{1}{\lambda_2} \! & \! \cdots \! & \! \frac{1}{\lambda_N} \! 
\end{array}
\right] 
\left\{ I_N - 
\left[
\begin{array}{ccc}
 V_{11} & \cdots & V_{1 N} \\
 \vdots &  \ddots & \vdots \\
 V_{N 1} & \cdots & V_{N N}\\
\end{array}
\right] \right\}^{-1}
\left[
\begin{array}{c}
 c_1 \left( \frac{2\lambda_1 - \varDelta}{2\lambda_1 + \varDelta} \right)^n 
	\mathrm{e}^{\frac{1}{\lambda_1}t_1+\lambda_1 t_2}\\
 \vdots \\
 c_N \left( \frac{2\lambda_N - \varDelta}{2\lambda_N + \varDelta} \right)^n 
	\mathrm{e}^{\frac{1}{\lambda_N}t_1+\lambda_N t_2}\\
\end{array}
\right].
\label{}
\end{align}
\end{subequations}

To reduce (\ref{soliton_t12}) to 
the $N$-soliton solution of 
the semi-discrete massive Thirring model (\ref{sd_MTMeqs}), 
we 
set 
\begin{equation}
t_1 = at, \hspace{5mm} t_2 = bt,
\label{t_12}
\end{equation}
which indeed implies \mbox{$\partial_t = a \partial_{t_1} + b \partial_{t_2}$}; 
note that 
the reduction condition \mbox{$\widebar{F}_n=F_n^\ast$}  
to realize 
the complex conjugation reduction 
(\ref{rq_vu}) 
requires 
the condition 
\mbox{$
 a, b \in \mathbb{R}$} 
(cf.~(\ref{FbarF})). 
Thus, 
the $N$-soliton solution of 
the semi-discrete massive Thirring model (\ref{sd_MTMeqs}) is given by 
\begin{subequations}
\label{soliton_t}
\begin{align}
u_n (t) & = \left[
\begin{array}{cccc}
\! 1 \! & \! 1 \! & \! \cdots \! & \! 1 \! 
\end{array}
\right] 
\left\{ I_N + 
\left[
\begin{array}{ccc}
 U_{11} & \cdots & U_{1 N} \\
 \vdots &  \ddots & \vdots \\
 U_{N 1} & \cdots & U_{N N}\\
\end{array}
\right] \right\}^{-1}
\left[
\begin{array}{c}
 c_1 \left( \frac{2\lambda_1 - \varDelta}{2\lambda_1 + \varDelta} \right)^n 
	\mathrm{e}^{\left(\frac{a}{\lambda_1}+b\lambda_1 \right) t} \\
\vdots \\
 c_N \left( \frac{2\lambda_N - \varDelta}{2\lambda_N + \varDelta} \right)^n 
	\mathrm{e}^{\left(\frac{a}{\lambda_N}+b\lambda_N \right) t} \\
\end{array}
\right], 
\label{}
\\[2mm]
q_n (t) & = -\mathrm{i} \left[
\begin{array}{cccc}
\! \frac{1}{\lambda_1} \! & \! \frac{1}{\lambda_2} \! & \! \cdots \! & \! \frac{1}{\lambda_N} \! 
\end{array}
\right] 
\left\{ I_N - 
\left[
\begin{array}{ccc}
 V_{11} & \cdots & V_{1 N} \\
 \vdots &  \ddots & \vdots \\
 V_{N 1} & \cdots & V_{N N}\\
\end{array}
\right] \right\}^{-1}
\left[
\begin{array}{c}
 c_1 \left( \frac{2\lambda_1 - \varDelta}{2\lambda_1 + \varDelta} \right)^n 
	\mathrm{e}^{\left(\frac{a}{\lambda_1}+b\lambda_1 \right) t}\\
 \vdots \\
 c_N \left( \frac{2\lambda_N - \varDelta}{2\lambda_N + \varDelta} \right)^n 
	\mathrm{e}^{\left(\frac{a}{\lambda_N}+b\lambda_N \right) t}\\
\end{array}
\right],
\label{}
\end{align}
where 
$U_{lj}$ and $V_{lj}$ are defined 
as 
(\ref{UV_lj}) with (\ref{t_12}), i.e.,
\begin{align}
& U_{lj} := \sum_{k=1}^N \frac{\mathrm{i} \lambda_j c_l c_k^\ast}
	{\left( \lambda_l + \lambda_k^\ast \right) \left( \lambda_k^\ast + \lambda_j \right)} 
	\left( \frac{2\lambda_l - \varDelta}{2\lambda_l + \varDelta} \right)^n 
\left( \frac{2\lambda_k^\ast - \varDelta}{2\lambda_k^\ast + \varDelta} \right)^n 
\mathrm{e}^{\left[ a \left( \frac{1}{\lambda_l} + \frac{1}{\lambda_k^\ast} \right) + b \left(\lambda_l +\lambda_k^\ast \right) \right] t}, 
\label{U_lj_t} \\[1mm]
& V_{lj} := \sum_{k=1}^N \frac{\mathrm{i} \lambda_k^\ast c_l c_k^\ast}
	{\left( \lambda_l + \lambda_k^\ast \right) \left( \lambda_k^\ast + \lambda_j \right)} 
	\left( \frac{2\lambda_l - \varDelta}{2\lambda_l + \varDelta} \right)^n 
\left( \frac{2\lambda_k^\ast - \varDelta}{2\lambda_k^\ast + \varDelta} \right)^n 
\mathrm{e}^{\left[ a \left( \frac{1}{\lambda_l} + \frac{1}{\lambda_k^\ast} \right) + b \left(\lambda_l +\lambda_k^\ast \right) \right] t}.
\label{V_lj_t}
\end{align}
\end{subequations}

In the simplest case 
\mbox{$N=1$}, 
we obtain the one-soliton solution of 
the semi-discrete massive Thirring model (\ref{sd_MTMeqs}) 
as 
\begin{align}
u_n (t) & = \frac{c_1 \left( \frac{2\lambda_1 - \varDelta}{2\lambda_1 + \varDelta} \right)^n 
	\mathrm{e}^{\left(\frac{a}{\lambda_1}+b\lambda_1 \right) t}}
{1+\frac{\mathrm{i} \lambda_1 c_1 c_1^\ast}
	{\left( \lambda_1 + \lambda_1^\ast \right)^2} 
	\left( \frac{2\lambda_1 - \varDelta}{2\lambda_1 + \varDelta} \right)^n 
\left( \frac{2\lambda_1^\ast - \varDelta}{2\lambda_1^\ast + \varDelta} \right)^n 
\mathrm{e}^{\left[ a \left( \frac{1}{\lambda_1} + \frac{1}{\lambda_1^\ast} \right) + b \left(\lambda_1 +\lambda_1^\ast \right) \right] t}}, 
\nonumber 
\\[2mm]
q_n (t) & = \frac{-\mathrm{i} \frac{c_1}{\lambda_1} 
  \left( \frac{2\lambda_1 - \varDelta}{2\lambda_1 + \varDelta} \right)^n 
	\mathrm{e}^{\left(\frac{a}{\lambda_1}+b\lambda_1 \right) t}}
{1-\frac{\mathrm{i} \lambda_1^\ast c_1 c_1^\ast}
	{\left( \lambda_1 + \lambda_1^\ast \right)^2} 
	\left( \frac{2\lambda_1 - \varDelta}{2\lambda_1 + \varDelta} \right)^n 
\left( \frac{2\lambda_1^\ast - \varDelta}{2\lambda_1^\ast + \varDelta} \right)^n 
\mathrm{e}^{\left[ a \left( \frac{1}{\lambda_1} + \frac{1}{\lambda_1^\ast} \right) + b \left(\lambda_1 +\lambda_1^\ast \right) \right] t}}.
\nonumber 
\end{align}

\section{Conclusions} 

In this paper, 
we started with
the 
semi-discrete Lax-pair representation 
(\ref{line}) with 
(\ref{sdMTM_L}) and (\ref{sdMTM_M}). 
The semi-discrete 
zero-curvature condition (\ref{Lax_eq00}) 
provides the differential-difference system (\ref{system00}); 
the auxiliary function $\Lambda_n$ admits the explicit expression 
(\ref{L_exp1}), 
which allows us to 
discard the last equation 
(\ref{Lambda_t}) in (\ref{system00}). 
If the parameters 
$\varDelta$, $a$ and $b$ 
in (\ref{system00}) 
are 
all 
real, 
we can impose the complex conjugation reduction 
(\ref{rq_vu})
to obtain 
the 
semi-discrete system (\ref{sd_MTMeqs}), 
which 
reduces in the continuous limit \mbox{$\varDelta \to 0$}  
to 
the massive Thirring model in 
non-characteristic and non-laboratory 
coordinates (\ref{cMTM0}). 
We can decompose the time part of the original Lax-pair representation (\ref{line_t}) with (\ref{sdMTM_M}) 
into 
two linear problems (\ref{U_t1}) and (\ref{U_t2}),  
where the decomposition means that the original time evolution is 
a linear combination of two commuting flows: 
\mbox{$\partial_t = a \partial_{t_1} + b \partial_{t_2}$}, 
\mbox{$[\partial_{t_1}, \partial_{t_2}]=0$}. 
This decomposition enables 
us to 
use 
the solution formulas for the massive Thirring model
based on the 
inverse scattering transform 
(see~Proposition~A.1 in \cite{TsuJMP11}) 
and 
presented in \cite{TsuJMP10}.  
By considering the 
reflectionless case and solving the linear integral equations,
we obtain the $N$-soliton solution (\ref{soliton_t}) of the semi-discrete massive Thirring model 
(\ref{sd_MTMeqs}). 


\appendix
\section{
The 
discrete spectral problem 
arising from a 
B\"acklund--Darboux 
transformation} 

The Lax-pair representation (\ref{MTM_t12})
for the 
massive Thirring model in light-cone 
coordinates (\ref{c-MTM1})
admits 
the binary 
B\"acklund--Darboux 
transformation~\cite{Holod78,
Nissimov79,Morris80, Pri81, 
David84}, 
which 
is compatible with the complex conjugation reduction (\ref{rq_vu}). 
With a particular choice of the B\"acklund parameters 
and an inessential overall factor, 
the binary B\"acklund--Darboux 
transformation
can be written in four 
equivalent ways 
as follows~\cite{Me2015,Me2024}: 
\begin{align}
\left[
\begin{array}{c}
 \Psi_{1, n+1}  \\
 \Psi_{2, n+1} \\
\end{array}
\right]
&= \left\{ 
\left[
\begin{array}{cc}
 -1 -\frac{\mathrm{i}\varDelta}{2} \zeta^2 & 0 \\ 
 0 & 1 - \frac{\mathrm{i}\varDelta}{2} \zeta^2 \\
\end{array}
\right] + 2\zeta
 \left[
\begin{array}{cc}
 \zeta &  \phi_n \\
 \chi_n &  -\frac{2\mathrm{i}}{\varDelta \zeta} \\
\end{array}
\right]^{-1} \right\}
\left[
\begin{array}{c}
 \Psi_{1,n}  \\
 \Psi_{2,n} \\
\end{array}
\right]
\nonumber \\[1mm]
&= \left\{ 
\left[
\begin{array}{cc}
 \frac{1+\frac{\mathrm{i}\varDelta}{2} \phi_n \chi_n}{1-\frac{\mathrm{i}\varDelta}{2} \phi_n \chi_n} & 0 \\ 
 0 & 1\\
\end{array}
\right] - \frac{\mathrm{i}\varDelta}{2}
 \left[
\begin{array}{cc}
 \zeta^2 & \zeta \frac{2\phi_n}{1-\frac{\mathrm{i}\varDelta}{2} \phi_n \chi_n} \\
 \zeta \frac{2\chi_n}{1-\frac{\mathrm{i}\varDelta}{2} \phi_n \chi_n} 
	&  -\zeta^2  \frac{1+\frac{\mathrm{i}\varDelta}{2} \phi_n \chi_n}{1-\frac{\mathrm{i}\varDelta}{2} \phi_n \chi_n} \\
\end{array}
\right] \right\}
\left[
\begin{array}{c}
 \Psi_{1,n}  \\
 \Psi_{2,n} \\
\end{array}
\right]
\nonumber \\[1mm]
& = \left[
\begin{array}{cc}
 \zeta & \phi_n \\ 
 \chi_n & \frac{2\mathrm{i}}{\varDelta \zeta} \\
\end{array}
\right]
 \left[
\begin{array}{cc}
 1-\frac{\mathrm{i}\varDelta}{2} \zeta^2 & 0 \\
 0 &  -1 - \frac{\mathrm{i}\varDelta}{2} \zeta^2 \\
\end{array}
\right] 
\left[
\begin{array}{cc}
 \zeta & \phi_n \\ 
 \chi_n & -\frac{2\mathrm{i}}{\varDelta \zeta} \\
\end{array}
\right]^{-1}
\left[
\begin{array}{c}
 \Psi_{1,n}  \\
 \Psi_{2,n} \\
\end{array}
\right]
\nonumber \\[1mm]
&= \left[
\begin{array}{cc}
 \zeta & \phi_n \\ 
 \chi_n & -\frac{2\mathrm{i}}{\varDelta \zeta} \\
\end{array}
\right]^{-1} 
 \left[
\begin{array}{cc}
 1 - \frac{\mathrm{i}\varDelta}{2} \zeta^2 & 0 \\
 0 &  -1 - \frac{\mathrm{i}\varDelta}{2} \zeta^2 \\
\end{array}
\right] 
\left[
\begin{array}{cc}
 \zeta & \phi_n \\ 
 \chi_n & \frac{2\mathrm{i}}{\varDelta \zeta} \\
\end{array}
\right] 
\left[
\begin{array}{c}
 \Psi_{1,n}  \\
 \Psi_{2,n} \\
\end{array}
\right],
\label{DBtra}
\end{align}
where 
$\phi_n$ and $\chi_n$ are defined in terms of the linear eigenfunction 
as 
\begin{equation}
\phi_n := -\left.  \frac{\zeta \Psi_{1,n}}{\Psi_{2,n}} \right|_{\zeta^2=\frac{2\mathrm{i}}{\varDelta}}, \hspace{5mm}
\chi_n := \left. \frac{\zeta \Psi_{2,n}}{\Psi_{1,n}} \right|_{\zeta^2=-\frac{2\mathrm{i}}{\varDelta}}. 
\label{intermediate_def}
\end{equation}
More precisely, we can use 
different 
(i.e., linearly independent) 
eigenfunctions 
to define $\phi_n$ and $\chi_n$. 

The compatibility condition of 
(\ref{U_t1}) and (\ref{DBtra}) is equivalent to 
the system of four equations: 
\begin{subequations}
\label{t1_BD}
\begin{align}
& \mathrm{i} \phi_{n,t_1} = \frac{2\mathrm{i}}{\varDelta}q_n - \frac{2\mathrm{i}}{\varDelta} \phi_n +q_n r_n \phi_n - r_n \phi_n^2, 
\label{t1_BD1}
\\[1mm]
& \mathrm{i} \chi_{n,t_1} = \frac{2\mathrm{i}}{\varDelta}r_n - \frac{2\mathrm{i}}{\varDelta} \chi_n - q_n r_n \chi_n + q_n \chi_n^2, 
\label{t1_BD2}
\\[1mm]
& q_{n+1} = - \frac{1 -\frac{\mathrm{i}\varDelta}{2} \phi_n \chi_n}{1 +\frac{\mathrm{i}\varDelta}{2} \phi_n \chi_n} q_n 
	+ \frac{2\phi_n}{1+\frac{\mathrm{i}\varDelta}{2} \phi_n \chi_n}, 
\label{t1_BD3}
\\[1mm]
& r_{n+1} = -\frac{1+\frac{\mathrm{i}\varDelta}{2} \phi_n \chi_n}{1-\frac{\mathrm{i}\varDelta}{2} \phi_n \chi_n} r_n 
	+ \frac{2\chi_n}{1 - \frac{\mathrm{i}\varDelta}{2} \phi_n \chi_n}. 
\label{t1_BD4}
\end{align}
\end{subequations}
Note that 
the 
first two relations (\ref{t1_BD1}) and (\ref{t1_BD2}) 
can also be derived from 
the definition of $\phi_n$ and $\chi_n$ 
in 
(\ref{intermediate_def}) 
and the time-evolution equation (\ref{U_t1}). 

The compatibility condition of 
(\ref{U_t2}) and (\ref{DBtra}) is equivalent to 
the system of four equations: 
\begin{subequations}
\label{t2_BD}
\begin{align}
& \mathrm{i} \phi_{n,t_2} = u_n -\frac{\mathrm{i}\varDelta}{2} \phi_n -u_n v_n \phi_n +\frac{\mathrm{i}\varDelta}{2} v_n \phi_n^2, 
\label{t2_BD1}
\\[1mm]
& \mathrm{i} \chi_{n,t_2} = -v_n -\frac{\mathrm{i}\varDelta}{2} \chi_n + u_n v_n \chi_n +\frac{\mathrm{i}\varDelta}{2} u_n \chi_n^2, 
\label{t2_BD2}
\\[1mm]
& u_{n+1} = \frac{1 +\frac{\mathrm{i}\varDelta}{2} \phi_n \chi_n}{1-\frac{\mathrm{i}\varDelta}{2} \phi_n \chi_n} u_n 
	- \frac{\mathrm{i}\varDelta \phi_n}{1-\frac{\mathrm{i}\varDelta}{2} \phi_n \chi_n}, 
\label{t2_BD3}
\\[1mm]
& v_{n+1} = \frac{1-\frac{\mathrm{i}\varDelta}{2} \phi_n \chi_n}{1+\frac{\mathrm{i}\varDelta}{2} \phi_n \chi_n} v_n 
	+ \frac{\mathrm{i}\varDelta \chi_n}{1+\frac{\mathrm{i}\varDelta}{2} \phi_n \chi_n}. 
\label{t2_BD4}
\end{align}
\end{subequations}
Note that 
the first two relations (\ref{t2_BD1}) and (\ref{t2_BD2}) 
can also be derived from the definition of $\phi_n$ and $\chi_n$ 
in 
(\ref{intermediate_def}) 
and the time-evolution equation (\ref{U_t2}). 

By introducing 
the 
auxiliary 
function 
$\Lambda_n$ as
\begin{equation}
\Lambda_n := \frac{1+\frac{\mathrm{i}\varDelta}{2} \phi_n \chi_n}{1-\frac{\mathrm{i}\varDelta}{2} \phi_n \chi_n}, 
\label{La_int}
\end{equation}
the relations (\ref{t1_BD3}), (\ref{t1_BD4}), (\ref{t2_BD3}) and (\ref{t2_BD4}) 
can be rewritten as 
\begin{subequations}
\label{qruv_re}
\begin{align}
& \Lambda_n q_{n+1} + q_n = \frac{2\phi_n}{1-\frac{\mathrm{i}\varDelta}{2} \phi_n \chi_n}, 
\label{La_pc1}
\\[1mm]
& r_{n+1} + \Lambda_n r_n = \frac{2\chi_n}{1-\frac{\mathrm{i}\varDelta}{2} \phi_n \chi_n}, 
\label{La_pc2}
\\[1mm]
& u_{n+1} -\Lambda_n u_n = -\frac{\mathrm{i}\varDelta \phi_n}{1-\frac{\mathrm{i}\varDelta}{2} \phi_n \chi_n}, 
\label{La_uv1}
\\[1mm]
& \Lambda_n v_{n+1} -v_n = \frac{\mathrm{i}\varDelta \chi_n}{1-\frac{\mathrm{i}\varDelta}{2} \phi_n \chi_n},
\label{La_uv2}
\end{align}
\end{subequations}
respectively. 
Using (\ref{La_pc1}) and (\ref{La_pc2}) with (\ref{La_int}), 
we obtain the quadratic equation for $\Lambda_n$: 
\begin{equation}
1 + \frac{\mathrm{i}\varDelta}{2} (\Lambda_n q_{n+1} + q_n) (r_{n+1} + \Lambda_n r_n) 
= \left( \frac{1+\frac{\mathrm{i}\varDelta}{2} \phi_n \chi_n}{1-\frac{\mathrm{i}\varDelta}{2} \phi_n \chi_n} \right)^2 = \Lambda_n^2, 
\nonumber 
\end{equation}
which coincides with 
(\ref{L_quadra}) (or (\ref{L_quad})). 
The definition 
(\ref{La_int}) 
implies that $\Lambda_n$ can be expanded as 
\mbox{$\Lambda_n  
= 1 + \mathcal{O}(\varDelta)$} 
for a small value of $\varDelta$, 
so we 
can conclude that 
$\Lambda_n$ 
admits the explicit expression 
(\ref{L_exp1})
(cf.~(\ref{L_exp2})). 

With the aid of (\ref{La_int}), (\ref{La_pc1}) and (\ref{La_pc2}), 
the second expression in (\ref{DBtra}) 
for the binary 
B\"acklund--Darboux 
transformation 
can be identified with 
the original discrete spectral problem (\ref{line_s}) with (\ref{sdMTM_L}). 
In essence, 
the system (\ref{system01}) 
can be obtained from (\ref{t1_BD}) and (\ref{La_int}) by eliminating $\phi_n$ and $\chi_n$, 
while the 
system (\ref{system02}) can be obtained 
from (\ref{t2_BD1}), (\ref{t2_BD2}),  (\ref{La_int}) and (\ref{qruv_re})
by eliminating $\phi_n$ and $\chi_n$. 

If $q_n$, $r_n$, $u_n$ and $v_n$ satisfy the 
equations of motion for the 
massive Thirring model 
(\ref{c-MTM1}), 
we can confirm 
using (\ref{t1_BD1}), (\ref{t1_BD2}), (\ref{t2_BD1}) and (\ref{t2_BD2}) 
the commutativity of the two time derivatives:\ 
\mbox{$[\partial_{t_1}, \partial_{t_2}] \phi_n= [\partial_{t_1}, \partial_{t_2}] \chi_n =0$}, 
so the two systems (\ref{system01}) and (\ref{system02}) 
are 
indeed 
compatible. 


\addcontentsline{toc}{section}{References}
\begin{thebibliography}{99}

\bibitem{Thirring58}
W.\ E.\ Thirring: 
{\em A soluble relativistic field theory}, 
Annals of Physics {\bf 3} (1958) 91--112.

\bibitem{NCQ83}
F.\ W.\ Nijhoff, H.\ W.\ Capel 
and G.\ R.\ W.\ Quispel: 
{\em Integrable lattice version of the massive Thirring model and 
its linearization}, Phys.\ Lett.\ A {\bf 98} (1983) 83--86. 

\bibitem{NCQL}
F.\ W.\ Nijhoff, H.\ W.\ Capel, G.\ R.\ W.\ Quispel and J.\ van der Linden: 
{\em The derivative nonlinear Schr\"odinger equation and 
the massive Thirring model}, 
Phys.\ Lett.\ A {\bf 93} (1983) 455--458.

\bibitem{Me2015}
T.\ Tsuchida: {\em On a new integrable discretization of 
the derivative nonlinear Schr\"odinger $(\mbox{Chen--Lee--Liu})$ equation}, 
arXiv:1501.01956 [nlin.SI] (2015).

\bibitem{Me2024}
T.\ Tsuchida: {\em On an integrable discretization of the massive Thirring model in light-cone coordinates and the associated Yang--Baxter map}, 
arXiv:2408.13913 [nlin.SI] (2024).


\bibitem{Peli_Joshi}
N.\ Joshi and D.\ E.\ Pelinovsky: 
{\em Integrable semi-discretization of the massive Thirring system in laboratory coordinates}, 
J.\ Phys.\ A:\ Math.\ Theor.\ {\bf 52} (2019) 03LT01.


\bibitem{Pelinovsky2019}
T.\ Xu and D.\ E.\ Pelinovsky: 
{\em Darboux transformation and soliton solutions of the semi-discrete massive Thirring model}, 
Phys.\ Lett.\ A {\bf 383} 
(2019) 125948.


\bibitem{AL1}
M.\ J.\ Ablowitz and J.\ F.\ Ladik: 
{\em Nonlinear differential--difference equations and Fourier analysis}, 
J.\ Math.\ Phys.\ {\bf 17} (1976) 1011--1018.

\bibitem{AL76}
M.\ J.\ Ablowitz 
and J.\ F.\ Ladik: 
{\em A nonlinear difference scheme and inverse scattering}, 
Stud.\ Appl.\ Math.\ {\bf 55} (1976) 213--229. 

\bibitem{CLL} 
H.\ H.\ Chen, Y.\ C.\ Lee and C.\ S.\ Liu: {\em Integrability of nonlinear
 Hamiltonian systems by inverse scattering method}, 
Phys.\ Scr.\ {\bf 20} (1979) 490--492.

\bibitem{Lax}
P.\ D.\ Lax: {\em Integrals of nonlinear equations of evolution and
  solitary waves}, Commun.\ Pure Appl.\ Math.\ {\bf 21} (1968) 467--490.

\bibitem{Kuz}
E.\ A.\ Kuznetsov and A.\ V.\ Mikhailov: 
{\em On the complete integrability of the two-dimensional classical Thirring model}, 
Theor.\ Math.\ Phys.\ {\bf 30} (1977) 193--200. 


\bibitem{Vek98}
V.\ E.\ Vekslerchik: 
{\em `Universality' of the Ablowitz--Ladik hierarchy}, 
arXiv:solv-int/9807005 (1998).

\bibitem{Vek02}
V.\ E.\ Vekslerchik: 
{\em Functional representation of the Ablowitz--Ladik hierarchy.\ II},
J.\ Nonlinear Math.\ Phys.\ {\bf 9} (2002) 157--180.





\bibitem{Getmanov87}
I.\ V.\ Barashenkov and B.\ S.\ Getmanov: 
{\em Multisoliton solutions in the scheme for unified description of integrable 
relativistic massive fields.\ Non-degenerate $sl(2, \mathbb{C})$ case}, 
Commun.\ Math.\ Phys.\ {\bf 112} (1987) 423--446.

\bibitem{Getmanov93}
I.\ V.\ Barashenkov and B.\ S.\ Getmanov: 
{\em The unified approach to integrable relativistic equations:\ 
Soliton solutions over nonvanishing backgrounds.\ II}, 
J.\ Math.\ Phys.\ {\bf 34} (1993) 3054--3072.

\bibitem{AL77}
M.\ J.\ Ablowitz and J.\ F.\ Ladik: 
{\em On the solution of a class of nonlinear partial difference equations}, 
Stud.\ Appl.\ Math.\ {\bf 57} (1977) 1--12. 

\bibitem{Chiu77}
S.-C.\ Chiu and J.\ F.\ Ladik: 
{\em Generating exactly soluble nonlinear discrete evolution
 equations by a generalized Wronskian technique}, 
J.\ Math.\ Phys.\ {\bf 18} (1977) 690--700.

\bibitem{2010JPA}
T.\ Tsuchida: 
{\em A systematic method for constructing time
discretizations of integrable lattice systems:\ local
equations of motion}, 
J.\ Phys.\ A:\ Math.\ Theor.\ {\bf 43} (2010) 415202; 
a longer 
version is arXiv:0906.3155 [nlin.SI].


\bibitem{Kako}
F.\ Kako and N.\ Mugibayashi: 
{\em Complete integrability of 
general nonlinear differential-difference equations 
solvable by the inverse method.\ II}, 
Prog.\ Theor.\ Phys.\ {\bf 61} (1979) 776--790.

\bibitem{Ize81}
A.\ G.\ Izergin and V.\ E.\ Korepin: 
{\em A lattice model related to the nonlinear Schr\"odinger equation}, 
Sov.\ Phys.\ Dokl.\ {\bf 26} (1981) 653--654.


\bibitem{KN2}
D.\ J.\ Kaup and A.\ C.\ Newell: 
{\em On the Coleman correspondence and the solution of the massive Thirring model}, 
Lett.\ Nuovo Cimento {\bf 20} (1977) 325--331.

\bibitem{Morris79}
H. C. Morris: {\em The massive Thirring model connection}, 
J.\ Phys.\ A:\ Math.\ Gen.\ {\bf 12} (1979) 131--134.

\bibitem{KaMoIno}
T.\ Kawata, T.\ Morishima and H.\ Inoue: 
{\em Inverse scattering method for the two-dimensional massive Thirring model}, 
J.\ Phys.\ Soc.\ Jpn.\ {\bf 47} (1979) 1327--1334.

\bibitem{GIK}
V.\ S.\ Gerdjikov, M.\ I.\ Ivanov and P.\ P.\ Kulish: 
{\em Quadratic bundle and 
nonlinear equations}, 
Theor.\ Math.\ Phys.\ {\bf 44} (1980) 784--795.




\bibitem{TsuJMP10}
T.\ Tsuchida: {\em New reductions of integrable matrix partial differential
equations:\ $Sp(m)$-invariant systems}, 
J.\ Math.\ Phys.\ {\bf 51} (2010) 053511.  

\bibitem{TsuJMP11}
T.\ Tsuchida: {\em Systematic method of generating new integrable systems
via inverse Miura maps}, J.\ Math.\ Phys.\ {\bf 52} (2011) 053503. 

\bibitem{Holod78}
P.\ I.\ Holod: {\em Pseudopotentials and Backlund transformation for Thirring equation}, 
preprint in Russian, Kiev, Report number:\ ITF-78-100P (1978) 14 pages.

\bibitem{Nissimov79}
E.\ R.\ Nissimov and S.\ J.\ Pacheva: 
{\em Backlund transformation in the classical massive Thirring model}, 
CR Acad.\ Sci.\ Bulg.\ {\bf 32} (1979) 1191--1194.

\bibitem{Morris80}
R.\ K.\ Dodd and H.\ C.\ Morris: 
{\em B\"acklund transformations}, 
Geometrical Approaches to Differential Equations 
(Lecture Notes in Math.\ {\bf 810}, 
Springer, Berlin, 
1980) pp.\ 63--94.

\bibitem{Pri81}
A.\ K.\ Prikarpatskii: {\em Geometrical structure and B\"acklund transformations 
of nonlinear evolution equations possessing a Lax representation}, 
Theor.\ Math.\ Phys.\ {\bf 46} (1981) 249--256.

\bibitem{David84}
D.\ David: {\em On an extension of the classical Thirring model}, 
J.\ Math.\ Phys.\ {\bf 25} (1984) 3424--3432. 

\end{thebibliography}
\end{document}